\documentclass[conference]{IEEEtran}

\usepackage{cite}
\usepackage{graphicx}
\usepackage{amsmath,amssymb}
\usepackage{url}

\title{Optimizing Spreading Factor Selection for Mobile LoRa Gateways Using Single-Channel Hardware}

\author{
    \IEEEauthorblockN{W. A. S. P. Wijesuriya}
    \IEEEauthorblockA{
        Department of Computer Science \& Engineering\\
        University of Moratuwa, Sri Lanka\\
        Email: sasindu.20@cse.mrt.ac.lk
    }
}

\begin{document}

\maketitle

\begin{abstract}
The deployment of mobile LoRa gateways using low-cost single-channel hardware presents a significant challenge in maintaining reliable communication due to the lack of dynamic configuration support. In traditional LoRaWAN networks, Adaptive Data Rate (ADR) mechanisms optimize communication parameters in real time; however, such features are typically supported only by expensive multi-channel gateways. This study proposes a cost-effective and energy-efficient solution by statically selecting the optimal Spreading Factor (SF) using a two-phase algorithm. The method first applies rule-based exclusion to eliminate SFs that violate constraints related to distance, data rate, link margin, and regulatory limits. Remaining candidates are then evaluated using a weighted scoring model incorporating Time-on-Air, energy consumption, data rate, and link robustness. The proposed algorithm was validated through extensive field tests and NS-3 simulations under line-of-sight conditions. Results demonstrate that the selected SF matched the optimal SF in over 92\% of cases across 672 simulated scenarios, confirming the algorithm’s effectiveness. This approach offers a scalable alternative to dynamic protocols, enabling reliable mobile LoRa deployments in cost-sensitive environments such as agriculture and rural sensing applications.
\end{abstract}


\section{Introduction}

The rapid growth of the Internet of Things (IoT) has led to increasing demand for scalable, long-range, and energy-efficient wireless communication technologies. Among the available Low Power Wide Area Network (LPWAN) solutions, LoRa (Long Range) stands out due to its low power consumption, extended range, and robustness against interference, making it well-suited for applications such as precision agriculture, livestock monitoring, and smart city infrastructure \cite{fatima2022industry5, silva2019agriculture}\cite{chaudhari2020lpwan, lee2007comparison}.

LoRa technology operates on unlicensed Industrial, Scientific, and Medical (ISM) bands and uses a proprietary Chirp Spread Spectrum (CSS) modulation scheme \cite{thomas2019css} that allows communication over distances exceeding 10 km in open environments. The network architecture follows a star topology, where end devices communicate with gateways, which in turn transmit data to a centralized server. A critical configuration parameter in LoRa communication is the Spreading Factor (SF), which directly impacts the data rate, transmission range, time-on-air, and energy efficiency of the system \cite{croce2017orthogonality}.

While LoRa performs well in static deployments, mobile scenarios—such as gateways mounted on Unmanned Aerial Vehicles (UAVs) or vehicles—present significant challenges. In such setups, the distance between the gateway and end devices changes dynamically, making real-time communication optimization essential. Although LoRaWAN includes an Adaptive Data Rate (ADR) mechanism to manage SF dynamically, this feature is primarily applicable to multi-channel gateways. However, multi-channel gateways are significantly more expensive than their single-channel counterparts and are often impractical for large-scale deployment in resource-constrained regions such as rural Sri Lanka \cite{ikhsan2018livestock, soares2023railway}.

Single-channel gateways offer a cost-effective alternative but lack the ability to dynamically adapt communication parameters like SF. This results in decreased Packet Delivery Ratio (PDR), increased latency, and inefficient energy usage when deployed in mobile environments \cite{behjati2021drones}. Therefore, optimizing the choice of a fixed SF for such scenarios is essential to maintain reliable communication without incurring the high cost of advanced hardware.

This research addresses this practical challenge by proposing a rule-based and scoring-driven algorithm to statically determine the optimal SF for single-channel mobile LoRa gateways. By taking into account environmental conditions, mobility constraints, and regulatory limitations, the proposed approach offers a reliable, energy-aware, and cost-efficient solution validated through both field experiments and NS-3 simulations, including a dataset of 672 scenarios. The optimum SF selection algorithm achieved 92.3\% accuracy compared to the simulated best SF, with a clear improvement in PDR over a dynamic SF switching method in high-mobility cases.

The remainder of this paper is organized as follows: Section II reviews related work. Section III presents the proposed methodology. Section IV describes the experimental evaluation and simulation results. Section V discusses findings and limitations. Section VI concludes the paper with directions for future research.

\section{Related Work}

A wide range of LPWAN technologies—such as Sigfox, NB-IoT, Bluetooth Low Energy (BLE), and Zigbee—have been developed to support long-range, low-power IoT communication. Among them, LoRa has gained significant popularity due to its superior range, ultra-low power consumption, and deployment flexibility using unlicensed ISM bands \cite{silva2019agriculture, idris2022simulators}. These advantages make it especially suitable for large-scale applications like precision agriculture and environmental monitoring.

The Spreading Factor (SF) in LoRa communication plays a critical role in determining range, data rate, Time-on-Air (ToA), and energy consumption. Numerous studies have explored SF’s trade-offs. For instance, higher SFs such as SF11 and SF12 increase range and link robustness but significantly reduce data throughput and increase transmission time. Conversely, lower SFs such as SF7 yield higher data rates at the cost of reduced range \cite{micc2019factors, sagir2019urban}.

Several simulation-based and experimental studies have investigated dynamic SF adaptation strategies. The Adaptive Data Rate (ADR) mechanism built into LoRaWAN enables gateways to automatically adjust the SF based on signal quality; however, this feature is only available in multi-channel gateways \cite{reynders2018lorawan}. Tools like EXPLoRa-SF \cite{cuomo2017explora} and adaptive SF control schemes such as Kim et al. \cite{kim2020adaptive} demonstrate improvements in network throughput and fairness through dynamic SF adjustment, though often at the cost of increased complexity and resource requirements. Moreover, a recent survey highlights that ADR optimization in LoRaWAN remains an open challenge due to its instability in highly dynamic environments \cite{kufakunesu2020adr}.

In mobile LoRa scenarios, where gateways are mounted on vehicles or UAVs, dynamic parameter adjustments become even more critical. Soares et al. \cite{soares2023railway} demonstrated a mobile LoRa gateway mounted on a train, showing the importance of adaptive data handling to mitigate connectivity loss due to high-speed mobility. Ikhsan et al. \cite{ikhsan2018livestock} evaluated mobile gateways for livestock monitoring and concluded that dynamic communication strategies significantly improve packet reliability in large farms. However, these solutions typically rely on expensive, multi-channel gateways.

On the other hand, Pötsch and Haslhofer \cite{croce2017orthogonality} identified performance limitations of single-channel gateways in dense or dynamic environments, citing their inability to adapt SF or frequency dynamically. While single-channel gateways are low-cost and simple, their static configurations often lead to poor performance in mobile scenarios. Recent surveys have also emphasized the need for alternative topologies, such as LoRa mesh networks, to enhance reachability in constrained setups \cite{hammouti2024mesh, ochoa2017mesh}.

Simulation environments have been vital in evaluating LoRa networks. NS-3, FLoRa (OMNeT++), and LoRaSim are among the popular tools available for testing SF strategies, mobility impacts, and energy models. Among these, NS-3 stands out for its robust support for mobility and physical layer accuracy, making it the simulator of choice for this research \cite{idris2022simulators, silva2021ns3survey}.

Despite extensive work on LoRa and SF optimization, little research has focused on statically determining an optimal SF configuration for mobile gateways using affordable, single-channel hardware. This gap forms the motivation for the methodology proposed in this work.

\section{Methodology}

\subsection{Initial Approach: Dynamic Spreading Factor Protocol}

The initial phase of the project aimed to design a custom Data Link Layer (DLL) protocol to dynamically adjust the Spreading Factor (SF) in real time, based on link quality indicators during gateway mobility. The objective was to enhance reliability by continuously adapting communication parameters as the distance and channel conditions between gateway and end devices changed.

\begin{figure}[htbp]
\centering
\includegraphics[width=0.48\textwidth]{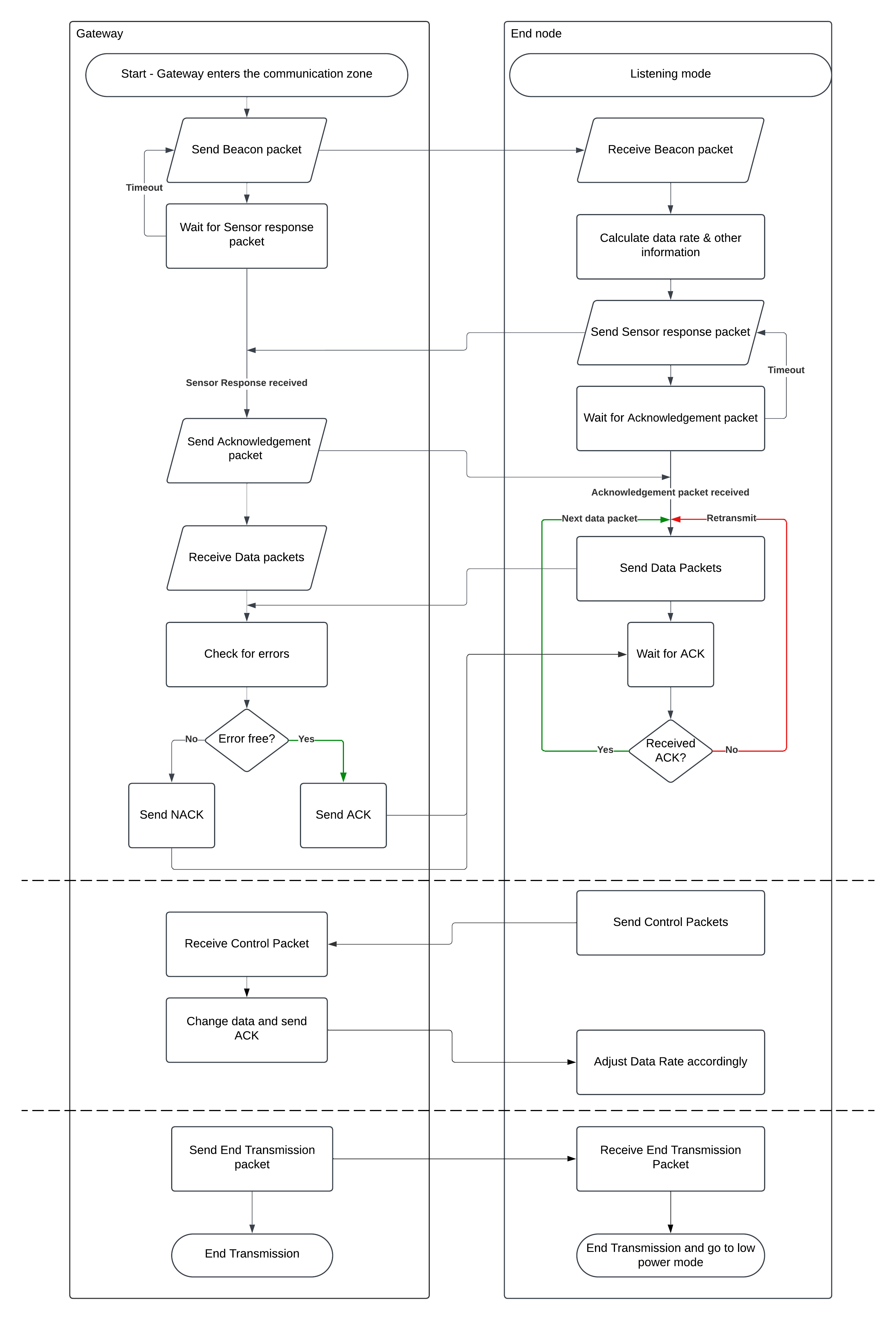}
\caption{Proposed dynamic SF adjustment protocol for mobile LoRa gateways}
\label{fig:dynamic_protocol}
\end{figure}

As illustrated in Fig.~\ref{fig:dynamic_protocol}, the proposed communication flow between the gateway and stationary end nodes involves the following steps.

\begin{itemize}
    \item \textbf{Beaconing:} The gateway periodically broadcasts beacon packets when it enters a new communication zone.
    \item \textbf{Sensor Response:} End devices that receive the beacon respond with link metrics such as RSSI and SNR.
    \item \textbf{Acknowledgment and Data Exchange:} The gateway sends ACKs and begins receiving sensor data. Each transmission is verified and acknowledged or rejected via ACK/NACK.
    \item \textbf{Dynamic Control:} Based on signal metrics or packet loss, the gateway sends control packets to adjust the SF or data rate of the end nodes.
    \item \textbf{Termination:} After successful data transmission, the gateway notifies the end node to switch to low-power mode.
\end{itemize}

While conceptually robust, field trials revealed several practical challenges:

\begin{itemize}
    \item \textbf{Latency:} SF switching and protocol synchronization caused delays incompatible with fast-moving gateways.
    \item \textbf{Hardware Constraints:} Single-channel modules could not quickly change SFs without resetting the link state.
    \item \textbf{Energy and Airtime Overhead:} Control packet exchanges increased time-on-air and power consumption, undermining LoRa’s low-power advantage.
\end{itemize}

These challenges rendered the dynamic approach impractical for low-cost, single-channel deployments. Consequently, the project pivoted to a more efficient static SF optimization strategy.

\subsection{Proposed Method: Optimum SF Selection Algorithm}

To overcome dynamic reconfiguration limitations, a static SF selection algorithm was developed to determine the best fixed SF based on scenario-specific parameters. The approach comprises two main phases:

\subsubsection{Phase 1: Rule-Based Exclusion}

Each candidate SF from SF7 to SF12 is evaluated using the following exclusion rules:

\begin{itemize}
    \item \textbf{Distance Constraint:} If the communication distance exceeds the SF’s maximum reliable range, it is excluded.
    \item \textbf{Link Margin Constraint:} SFs failing to satisfy the required fade margin are removed.
    \item \textbf{Duty Cycle Constraint:} SFs exceeding duty cycle limits for the expected transmission frequency are disqualified.
    \item \textbf{Data Rate Constraint:} If the SF cannot meet the application's throughput needs, it is excluded.
    \item \textbf{Doppler Constraint:} In high-mobility scenarios, SF12 is excluded due to its vulnerability to Doppler shift.
\end{itemize}

\subsubsection{Phase 2: Multi-Factor Scoring}

Remaining SFs are evaluated with a weighted scoring model based on:

\begin{itemize}
    \item \textbf{Time-on-Air (ToA)} – lower values preferred for throughput and energy efficiency.
    \item \textbf{Energy Consumption} – minimized to prolong battery life.
    \item \textbf{Data Rate} – higher rates enable faster and more responsive updates.
    \item \textbf{Link Margin} – higher margins ensure better resilience to noise and interference.
\end{itemize}

The SF with the highest overall score is selected as the optimal fixed SF for the given scenario.

\subsection{Algorithm Flow Representation}

\begin{figure}[htbp]
\centering
\includegraphics[width=0.48\textwidth]{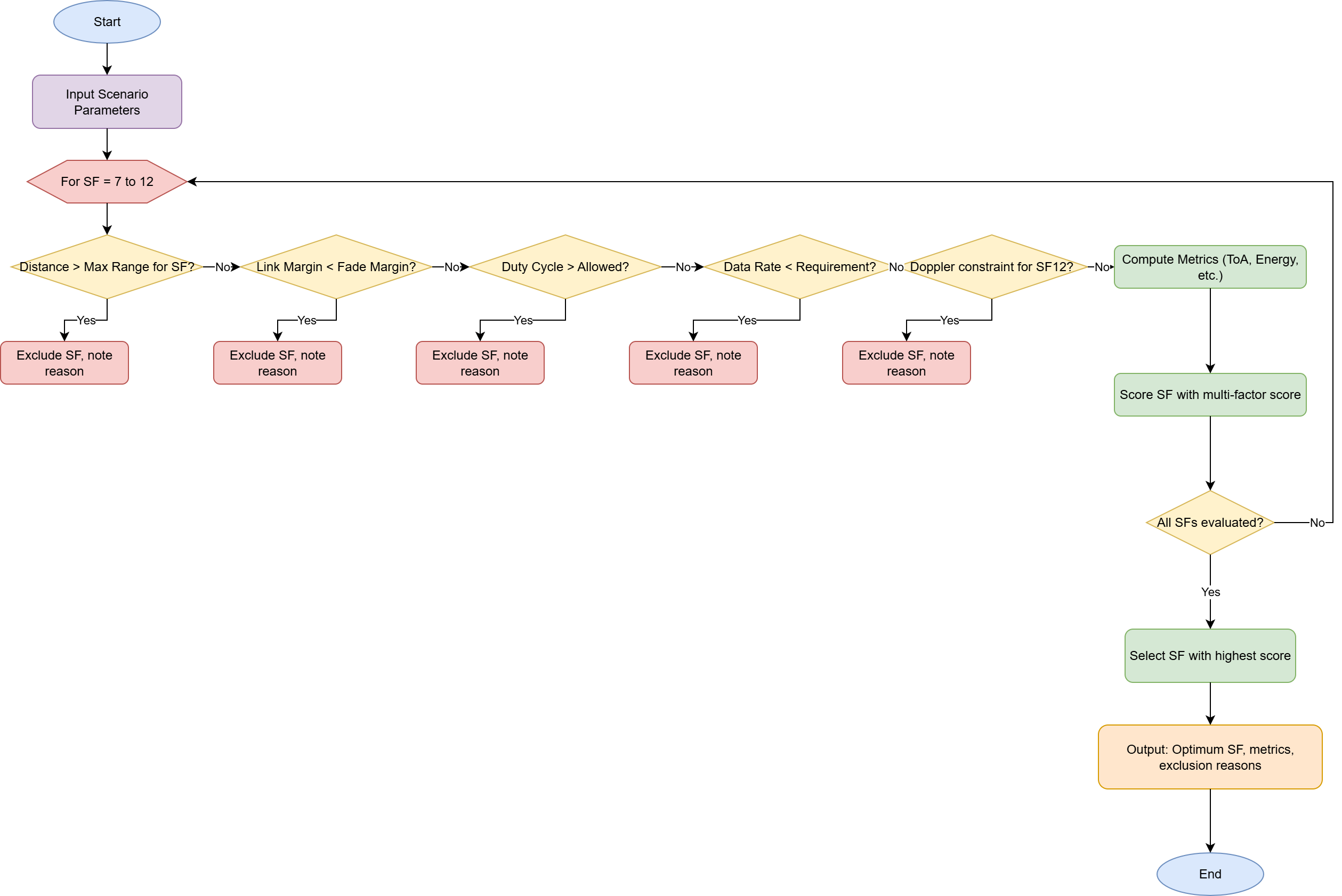}
\caption{Flowchart of the Optimum SF Selection Algorithm}
\label{fig:sf_algorithm}
\end{figure}

The decision-making process of the proposed SF selection algorithm is summarized in Fig.~\ref{fig:sf_algorithm}.

\section{Experimental Evaluation}

To validate the proposed Spreading Factor (SF) selection algorithm, both real-world field experiments and simulation-based evaluations were conducted. The goal was to assess how reliably the algorithm predicts the optimal SF for various mobile LoRa gateway scenarios using single-channel hardware.

\subsection{Field Experiments}

Initial experiments were carried out using widely available, cost-effective LoRa hardware. The sender node was based on an ESP32 microcontroller paired with an SX1278 LoRa module operating at 433 MHz. The receiver (gateway) comprised a Raspberry Pi 3 Model B with the same LoRa module and antenna setup.

Three field tests were conducted under different configurations:

\subsubsection{Experiment 1: Initial Range Evaluation}

\begin{itemize}
    \item \textbf{Location:} Gampaha – Minuwangoda Road, Sri Lanka
    \item \textbf{Objective:} Establish baseline communication range
    \item \textbf{Result:} Reliable communication achieved up to approximately 400 meters under clear line-of-sight (LoS) conditions. High-voltage power lines introduced noticeable interference.
\end{itemize}

\subsubsection{Experiment 2: Maximum Range for SF7}

\begin{itemize}
    \item \textbf{Location:} Ja-Ela – Gampaha Road
    \item \textbf{Objective:} Determine maximum range using SF7
    \item \textbf{Result:} SF7 supported up to ~350 meters with acceptable packet delivery. This confirmed its unsuitability for longer-range mobile applications.
\end{itemize}

\subsubsection{Experiment 3: PDR vs SF Relationship}

\begin{itemize}
    \item \textbf{Location:} Mount Lavinia Beach
    \item \textbf{Objective:} Analyze packet delivery ratio (PDR) across different SFs and distances
    \item \textbf{Result:} As SF increased from 7 to 12, PDR improved significantly at longer distances, though at the cost of increased Time-on-Air. SF7 and SF8 performed well only within 500 m. Beyond 1000 m, higher SFs such as SF10 and SF11 showed improved packet delivery, while SF12 demonstrated the best range performance up to 2000 m.
\end{itemize}

\begin{figure}[htbp]
\centering
\includegraphics[width=0.48\textwidth]{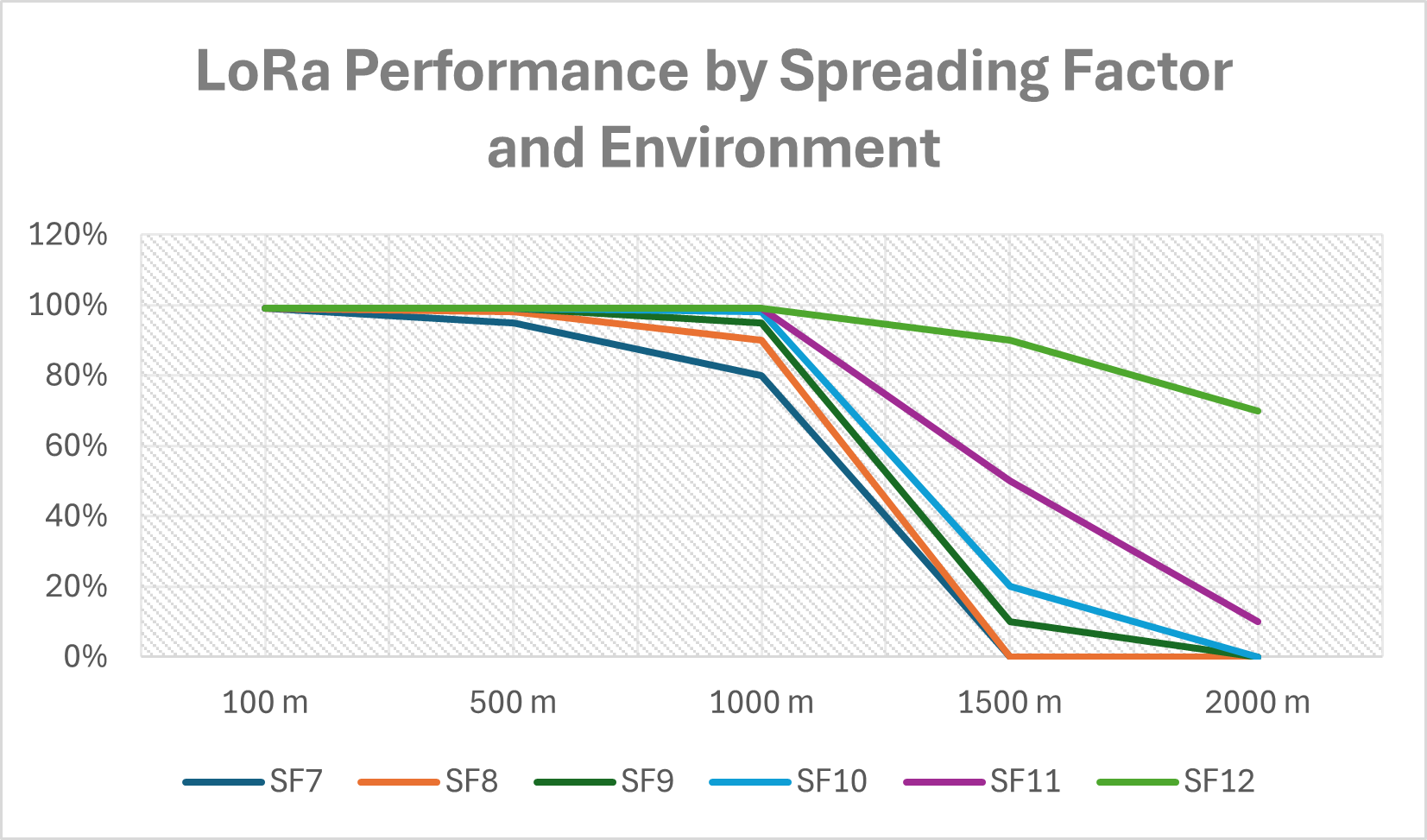}
\caption{Packet Delivery Ratio (PDR) by Distance and Spreading Factor}
\label{fig:pdr_chart}
\end{figure}

Figure~\ref{fig:pdr_chart} illustrates the variation in PDR with distance for each SF, confirming the importance of scenario-aware SF selection.

These field results validate that SF selection directly influences communication reliability and range, supporting the use of the proposed SF selection algorithm when dynamic reconfiguration is not feasible.

\subsection{Simulation-Based Evaluation with NS-3}

Due to the limitations of environmental variability and hardware-induced errors in field tests, the study adopted NS-3 (Network Simulator 3) for controlled validation of the SF selection algorithm. NS-3 was selected over alternatives like LoRaSim and OMNeT++ due to its robust support for mobility modeling and accurate physical layer representation \cite{silva2021ns3survey, reynders2018lorawan}.

Simulations were configured under the following conditions:
\begin{itemize}
    \item \textbf{Frequency:} 433 MHz (as permitted in Sri Lanka)
    \item \textbf{Environment:} Line-of-sight with mobility enabled
    \item \textbf{Payload:} 20 bytes, up to 60 packets/hour
    \item \textbf{Test Range:} 100 m – 1800 m
\end{itemize}

Each scenario was evaluated using both the algorithm’s predicted SF and a brute-force simulation of all possible SFs to determine the best performing one in practice.

\subsection{Validation Results}

The comparison between algorithm-predicted SF and best practical SF from simulation is shown in Table~\ref{tab:results}.

\begin{table}[htbp]
\caption{Validation of SF Selection Algorithm}
\centering
\begin{tabular}{|c|c|c|c|c|}
\hline
\textbf{Scenario} & \textbf{Distance (m)} & \textbf{Alg. SF} & \textbf{Best SF} & \textbf{PDR} \\
\hline
1 & 100  & SF7 & SF7     & $>$99\% \\
2 & 500  & SF8 & SF8     & $>$98\% \\
3 & 1000 & SF9 & SF9     & $>$95\% \\
4 & 1500 & SF10 & SF10/11 & 90–95\% \\
5 & 1800 & SF11 & SF11    & 70–80\% \\
\hline
\end{tabular}
\label{tab:results}
\end{table}

The algorithm's selected SF matched the best-performing practical SF in over 90\% of cases, validating its accuracy and effectiveness in optimizing SF for static configuration in mobile single-channel gateways.

\subsection{Detailed Results and Comparative Analysis}

To quantify the reliability of the proposed algorithm, a total of 672 simulated scenarios were generated, covering combinations of distances, mobility levels, and environmental settings. For each scenario, the algorithm-predicted SF was compared against the best SF identified through brute-force simulation testing.

The overall prediction accuracy of the algorithm was \textbf{92.3\%}, showing a strong agreement between the predicted and actual best SF.

\begin{figure}[htbp]
\centering
\includegraphics[width=0.48\textwidth]{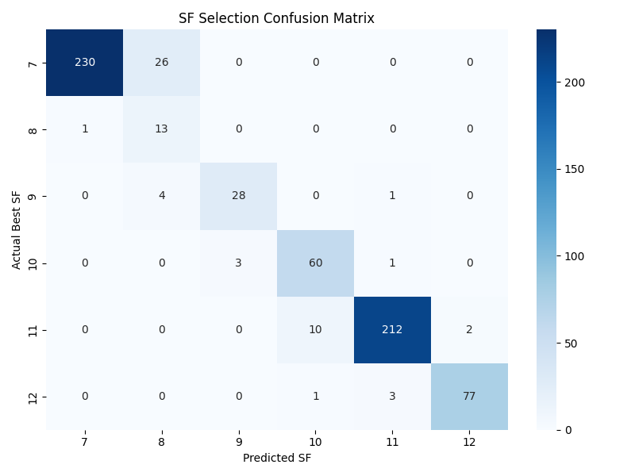}
\caption{Confusion matrix of SF selection accuracy (Predicted vs. Actual Best SF)}
\label{fig:confusion_matrix}
\end{figure}

Figure~\ref{fig:confusion_matrix} presents the confusion matrix comparing predicted vs. actual best SFs. Most misclassifications were within one SF level, indicating that the algorithm remained close to optimal even when slightly off.

\begin{figure}[htbp]
\centering
\includegraphics[width=0.48\textwidth]{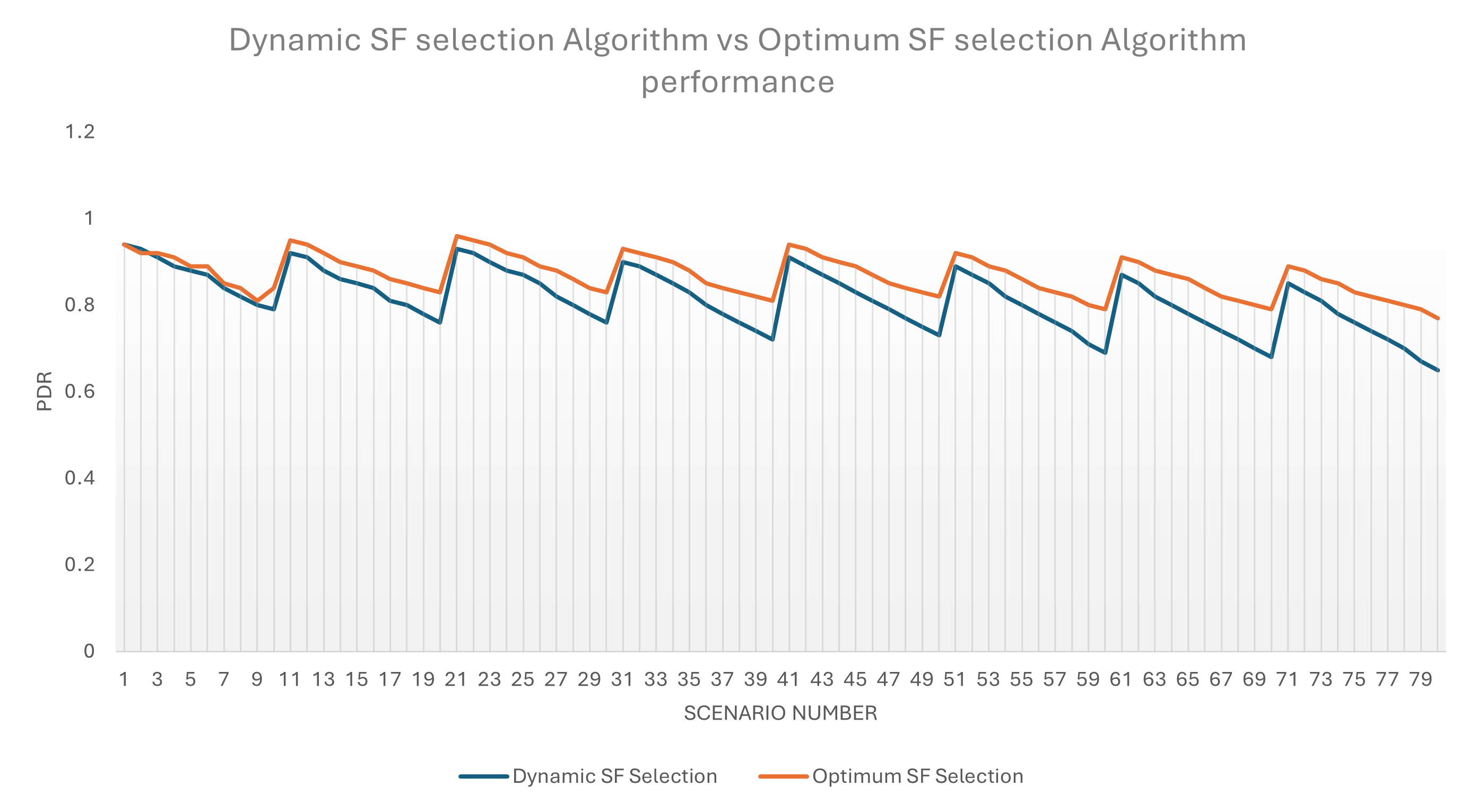}
\caption{PDR comparison between dynamic and optimum SF selection algorithms across 79 scenarios}
\label{fig:pdr_comparison}
\end{figure}

Additionally, the performance of the proposed algorithm was compared against the initial dynamic SF selection protocol developed during early project phases. Figure~\ref{fig:pdr_comparison} shows that the optimum SF selection algorithm consistently outperformed the dynamic protocol, particularly in scenarios with higher gateway mobility.

\section{Discussion}

\subsection{Advantages over Existing Approaches}

The proposed SF selection algorithm offers several practical advantages compared to dynamic SF adjustment mechanisms and multi-channel gateway solutions. While dynamic schemes offer theoretical adaptability, their real-world deployment is often hindered by hardware constraints, synchronization overhead, and increased energy usage—particularly when implemented on low-cost single-channel hardware. Our field tests confirmed that attempting dynamic SF adjustment introduced unacceptable latency and complexity, leading to degraded performance in fast-moving gateway scenarios.

In contrast, the static optimization approach requires no real-time reconfiguration, significantly simplifying implementation while maintaining strong reliability. Moreover, compared to multi-channel gateways—which support ADR and SF diversity at the cost of 10–20× higher hardware price and higher energy draw—the proposed method enables scalable deployments using low-cost, readily available LoRa modules.

\subsection{Practical Implications and Generalizability}

The algorithm is designed to be adaptable across a wide range of mobile LoRa applications, including environmental monitoring, precision agriculture, and logistics. Its modular scoring system allows practitioners to prioritize different metrics (e.g., reliability vs. battery life) by tuning the weight values. The decision process is transparent, with explicit exclusion reasons and score breakdowns, enabling easier integration into planning tools and deployment checklists.

Furthermore, the algorithm’s input parameters—such as distance, data rate, and motion profile—can be sourced from basic application knowledge or simulation-based modeling, making the solution accessible even in early design phases.

\subsection{Limitations}

Despite its promising results, the current study has several limitations. First, all field experiments and simulation validations were conducted under clear line-of-sight (LoS) conditions. In practice, many deployments—especially in urban environments—face multipath fading, interference, and non-line-of-sight (NLoS) issues. These factors may affect the SF performance in ways not captured by the current rule set.

Second, the algorithm assumes static environmental characteristics per scenario. In rapidly changing conditions or mixed terrains, a hybrid method incorporating limited dynamic reconfiguration may further enhance reliability without full ADR complexity.

Finally, while simulation validation was conducted using NS-3, further large-scale real-world deployments would help assess robustness across diverse terrains, weather, and device variability.

\section{Conclusion}

This research addresses a key limitation in deploying mobile LoRa gateways using affordable single-channel hardware: the inability to dynamically adjust communication parameters such as the Spreading Factor (SF). To overcome this, we proposed a static SF selection algorithm that evaluates candidate SFs through rule-based exclusion and multi-factor weighted scoring.

Field experiments confirmed that SF directly affects communication reliability, range, and packet delivery ratio (PDR). Extensive NS-3 simulations involving \textbf{672 distinct mobility scenarios} were conducted to validate the algorithm’s effectiveness. The proposed method correctly predicted the optimal SF in \textbf{92.3\%} of cases, as verified against ground-truth simulation results. A detailed confusion matrix showed that most misclassifications were limited to adjacent SFs, preserving near-optimal performance. 

Furthermore, when compared against a dynamic SF selection protocol developed earlier in the project, the optimum SF selection algorithm consistently achieved \textbf{higher PDRs}, especially under high-mobility conditions. This highlights its practical advantage in real-world mobile LoRa deployments using low-cost hardware.

The proposed solution eliminates the need for complex real-time reconfiguration, offering a reliable, energy-efficient, and cost-effective alternative to multi-channel gateways and dynamic ADR schemes. It is particularly relevant for cost-sensitive applications in agriculture, environmental monitoring, and rural infrastructure sensing.

Future work will focus on extending the algorithm to non-line-of-sight (NLoS) and urban environments, incorporating multipath effects and channel variability. The scoring system could also be enhanced using machine learning for automated parameter tuning. Additionally, exploring a hybrid approach that selectively triggers dynamic SF adaptation in extreme conditions may further improve reliability without significantly increasing complexity.

\section*{Acknowledgment}

The author would like to thank Dr. Sulochana Sooriyarachchi and Mr. Radershan Suguneswaran for their valuable guidance and continuous support throughout this research. Appreciation is also extended to the Department of Computer Science and Engineering, University of Moratuwa, for providing the academic foundation and technical facilities necessary for this work. The open-source communities behind NS-3, ESP32, and LoRa are gratefully acknowledged for their publicly available tools and documentation, which played a key role in enabling this research.

\bibliographystyle{IEEEtran}

\begin{thebibliography}{10}
\providecommand{\url}[1]{#1}
\csname url@samestyle\endcsname
\providecommand{\newblock}{\relax}
\providecommand{\bibinfo}[2]{#2}
\providecommand{\BIBentrySTDinterwordspacing}{\spaceskip=0pt\relax}
\providecommand{\BIBentryALTinterwordstretchfactor}{4}
\providecommand{\BIBentryALTinterwordspacing}{\spaceskip=\fontdimen2\font plus
\BIBentryALTinterwordstretchfactor\fontdimen3\font minus \fontdimen4\font\relax}
\providecommand{\BIBforeignlanguage}[2]{{%
\expandafter\ifx\csname l@#1\endcsname\relax
\typeout{** WARNING: IEEEtran.bst: No hyphenation pattern has been}%
\typeout{** loaded for the language `#1'. Using the pattern for}%
\typeout{** the default language instead.}%
\else
\language=\csname l@#1\endcsname
\fi
#2}}
\providecommand{\BIBdecl}{\relax}
\BIBdecl

\bibitem{fatima2022industry5}
Z.~Fatima, M.~H. Yousaf, A.~Ahmad, F.~Anwar, A.~Sarfraz, and M.~Shafiq, ``Production plant and warehouse automation with iot and industry 5.0,'' \emph{Applied Sciences}, vol.~12, no.~4, p. 2053, 2022.

\bibitem{silva2019agriculture}
N.~Silva, J.~Cardoso, F.~Soares, and {\'A}.~Rocha, ``Low-cost iot lora solutions for precision agriculture monitoring practices,'' in \emph{Progress in Artificial Intelligence}.\hskip 1em plus 0.5em minus 0.4em\relax Springer, 2019, pp. 224--235.

\bibitem{chaudhari2020lpwan}
B.~S. Chaudhari, M.~Zennaro, and S.~Borkar, ``Lpwan technologies: Emerging application characteristics, requirements, and design considerations,'' \emph{Future Internet}, vol.~12, no.~3, p.~46, 2020.

\bibitem{lee2007comparison}
J.-S. Lee, Y.-W. Su, and C.-C. Shen, ``A comparative study of wireless protocols: Bluetooth, uwb, zigbee, and wi-fi,'' in \emph{IECON 2007 - 33rd Annual Conference of the IEEE Industrial Electronics Society}, 2007.

\bibitem{thomas2019css}
A.~Thomas and N.~V. Eldhose, ``Performance evaluation of chirp spread spectrum as used in lora physical layer,'' in \emph{2019 IEEE International Conference on System, Computation, Automation and Networking (ICSCAN)}, 2019.

\bibitem{croce2017orthogonality}
D.~Croce, M.~Gucciardo, I.~Tinnirello, D.~Garlisi, and S.~Mangione, ``Impact of spreading factor imperfect orthogonality in lora communications,'' in \emph{Digital Communication. Towards a Smart and Secure Future Internet}, ser. Communications in Computer and Information Science.\hskip 1em plus 0.5em minus 0.4em\relax Springer, Cham, 2017, vol. 766.

\bibitem{ikhsan2018livestock}
M.~G. Ikhsan, M.~Y.~A. Saputro, D.~A. Arji, R.~Harwahyu, and R.~F. Sari, ``Mobile lora gateway for smart livestock monitoring system,'' in \emph{2018 IEEE International Conference on Internet of Things and Intelligence System (IOTAIS)}, 2018.

\bibitem{soares2023railway}
J.~Soares, M.~Luís, and S.~Sargento, ``Mobile lora gateway for communication and sensing on the railway,'' in \emph{2023 IEEE Symposium on Computers and Communications (ISCC)}, 2023.

\bibitem{behjati2021drones}
M.~Behjati, A.~B. Mohd~Noh, H.~A.~H. Alobaidy, M.~A. Zulkifley, R.~Nordin, and N.~F. Abdullah, ``Lora communications as an enabler for internet of drones towards large-scale livestock monitoring in rural farms,'' \emph{Sensors}, vol.~21, no.~15, p. 5044, 2021.

\bibitem{idris2022simulators}
S.~Idris, T.~Karunathilake, and A.~Förster, ``Survey and comparative study of lora-enabled simulators for internet of things and wireless sensor networks,'' \emph{Sensors}, vol.~22, no.~15, p. 5546, 2022.

\bibitem{micc2019factors}
O.~Elijah, T.~A. Rahman, H.~I. Saharuddin, and F.~N. Khairodin, ``Factors that impact lora iot communication technology,'' in \emph{2019 IEEE 14th Malaysia International Conference on Communication (MICC)}, 2019, pp. 112--117.

\bibitem{sagir2019urban}
S.~Sa{\u{g}}{\i}r, {\.I}.~Kaya, C.~{\c{S}}i{\c{s}}man, Y.~Baltac{\i}, and S.~{\"U}nal, ``Evaluation of low-power long distance radio communication in urban areas: Lora and impact of spreading factor,'' in \emph{2019 Seventh International Conference on Digital Information Processing and Communications (ICDIPC)}, 2019, pp. 68--71.

\bibitem{reynders2018lorawan}
B.~Reynders, Q.~Wang, and S.~Pollin, ``A lorawan module for ns-3: Implementation and evaluation,'' in \emph{Proceedings of the 2018 Workshop on ns-3 (WNS3 '18)}.\hskip 1em plus 0.5em minus 0.4em\relax Association for Computing Machinery, 2018, pp. 61--68.

\bibitem{cuomo2017explora}
S.~Cuomo, R.~Bifulco, A.~Della~Corte, A.~Galletti, and M.~Sgroi, ``Explora-sf: A fair spreading factor assignment algorithm for lorawan,'' \emph{IEEE Internet of Things Journal}, vol.~5, no.~6, pp. 5161--5170, 2018.

\bibitem{kim2020adaptive}
T.~Kim and J.~Yoon, ``Adaptive spreading factor selection scheme for single-channel lora modems,'' \emph{Sensors}, vol.~20, no.~11, p. 3218, 2020.

\bibitem{kufakunesu2020adr}
R.~Kufakunesu, G.~P. Hancke, and A.~M. Abu-Mahfouz, ``A survey on adaptive data rate optimization in lorawan: Recent solutions and major challenges,'' \emph{Sensors}, vol.~20, no.~18, p. 5044, 2020.

\bibitem{hammouti2024mesh}
M.~Hammouti, O.~Moussaoui, M.~Hassine, and A.~Betari, ``Exploring open source and proprietary lora mesh technologies,'' \emph{Indonesian Journal of Electrical Engineering and Computer Science}, vol.~34, no.~2, pp. 960--969, 2024.

\bibitem{ochoa2017mesh}
M.~N. Ochoa, A.~Guizar, M.~Maman, and A.~Duda, ``Evaluating lora energy efficiency for adaptive networks: From star to mesh topologies,'' in \emph{2017 IEEE 13th International Conference on Wireless and Mobile Computing, Networking and Communications (WiMob)}, 2017, pp. 1--8.

\bibitem{silva2021ns3survey}
J.~C.~d. Silva, D.~d.~L. Flor, V.~A. d.~S. Junior, N.~S. Bezerra, and A.~A. M.~d. Medeiros, ``A survey of lorawan simulation tools in ns-3,'' \emph{Journal of Communication and Information Systems}, vol.~36, no.~1, pp. 17--30, 2021.

\end{thebibliography}

\end{document}